\documentclass[lettersize,journal]{IEEEtran}
\usepackage{amsmath,amsfonts}
\usepackage{algorithmic}
\usepackage{array}
\usepackage[caption=false,font=normalsize,labelfont=sf,textfont=sf]{subfig}
\usepackage{textcomp}
\usepackage{stfloats}
\usepackage{url}
\usepackage{verbatim}
\usepackage{graphicx}
\def\BibTeX{{\rm B\kern-.05em{\sc i\kern-.025em b}\kern-.08em
    T\kern-.1667em\lower.7ex\hbox{E}\kern-.125emX}}
\usepackage{balance}

\usepackage{booktabs} % For formal tables

\usepackage{amsmath}
\usepackage{amssymb}
\usepackage{graphicx}
\usepackage{multirow}
\usepackage{enumitem}
\usepackage{makecell}

\usepackage[font={bf}]{caption}
\usepackage[ruled,linesnumbered,boxed]{algorithm2e}
\usepackage{color, url}
\usepackage{xspace} 
\usepackage{balance}

\newcommand{\ie}{\text{i.e., }}         % that is
\newcommand{\eg}{\text{e.g., }}         % for example
\newcommand{\etal}{\text{et al}}        % and others
\usepackage{threeparttable}
\usepackage{colortbl}       % for table
\usepackage{makecell}       % for table
\usepackage{tabularx}       % for table
\usepackage{bigstrut}       % for bigstrut
\usepackage{multirow}       % for multirow
\usepackage{colortbl}       % for cell color
\usepackage{ulem}[normalem] % for underline

\usepackage{tabu}
\usepackage{commath}

\usepackage[table]{xcolor}

\usepackage{ctable}
\usepackage{array}
\usepackage{arydshln}

\newcolumntype{L}[1]{>{\raggedright\let\newline\\\arraybackslash\hspace{0pt}}m{#1}}
\newcolumntype{C}[1]{>{\centering\let\newline\\\arraybackslash\hspace{0pt}}m{#1}}
\newcolumntype{R}[1]{>{\raggedleft\let\newline\\\arraybackslash\hspace{0pt}}m{#1}}

\renewcommand{\emph}[1]{\textit{#1}}

\begin{document}
\title{Correct and Weight: A Simple Yet Effective Loss for  Implicit Feedback Recommendation}

%~\IEEEmembership{Member,~IEEE,}
\author{Minglei Yin,  Chuanbo Hu, Bin Liu, Neil Zhenqiang Gong, Yanfang(Fanny) Ye, and Xin Li,~\IEEEmembership{Fellow,~IEEE}
%~\IEEEmembership{Member,~IEEE,}
\IEEEcompsocitemizethanks{
\IEEEcompsocthanksitem Minglei Yin and Xin Li are with the Department of Computer Science,  University at Albany,  Albany, NY 12222; E-mail:  myin@albany.edu, xli48@albany.edu
\protect\\
\IEEEcompsocthanksitem Bin Liu is with the Department of Management Information Systems, West Virginia University, 1601 University Avenue, 
Morgantown, WV 26506; E-mail: bin.liu1@mail.wvu.edu; 
\protect\\
\IEEEcompsocthanksitem Neil Zhenqiang Gong is with the Department of Electrical and Computer Engineering, Duke University, Box 90291
Durham, NC 27708; E-mail: neil.gong@duke.edu
\protect\\
\IEEEcompsocthanksitem Yanfang(Fanny) Ye is with the Department of Computer Science and Engineering, University of Notre Dame, Notre Dame, IN 46556; E-mail: yye7@nd.edu
% note need leading \protect in front of \\ to get a newline within \thanks as
% \\ is fragile and will error, could use \hfil\break instead.
}% <-this % stops an unwanted space
%\thanks{Manuscript received June xx, 2022}
\thanks{This work is partially supported by the NSF under grants CMMI-2146076, CNS-2125958, CCSS-2348046, and CNS-2125977, and by WVHEPC RCG23-007.}
}

\markboth{Journal of \LaTeX\ Class Files,~Vol.~18, No.~9, September~2020}%
{How to Use the IEEEtran \LaTeX \ Templates}

\maketitle

\begin{abstract}
Learning from implicit feedback has become the standard paradigm for modern recommender systems. However, this setting is fraught with the persistent challenge of false negatives, where unobserved user-item interactions are not necessarily indicative of negative preference. This ambiguity, often exacerbated by exposure bias, can severely degrade recommendation performance by providing misleading training signals. To address this critical issue, this paper introduces a novel and principled loss function, named \textbf{Corrected and Weighted (CW)} loss, that systematically corrects for the impact of false negatives within the training objective. Our approach integrates two key techniques. First, inspired by Positive-Unlabeled (PU) learning, we debias the negative sampling process by re-calibrating the assumed negative distribution. By theoretically approximating the true negative distribution ($p^-$) using the observable general data distribution ($p$) and the positive interaction distribution ($p^+$), our method provides a more accurate estimate of the likelihood that a sampled unlabeled item is truly negative. Second, we introduce a dynamic re-weighting mechanism that modulates the importance of each negative instance based on the model's current prediction. This scheme encourages the model to enforce a larger ranking margin between positive items and confidently predicted (i.e., easy) negative items, while simultaneously down-weighting the penalty on uncertain negatives that have a higher probability of being false negatives. A key advantage of our approach is its elegance and efficiency; it requires no complex modifications to the data sampling process or significant computational overhead, making it readily applicable to a wide array of existing recommendation models. Extensive experiments conducted on four large-scale, sparse benchmark datasets demonstrate the superiority of our proposed loss. The results show that our method consistently and significantly outperforms a suite of state-of-the-art loss functions across multiple ranking-oriented metrics. This work underscores the importance of directly addressing the false negative problem in the loss design and offers a simple yet powerful solution.
\end{abstract}


%%%%%%%%%%%%%%%%%%%%%%%%%%%%%%%%%%%%%%%%%%%%%%
\section{Introduction}\label{sec:intro}
Modern recommender systems have become indispensable components in large-scale content platforms, routing users to items of potential interest while mitigating information overload via mining historical user-item interaction data. 
\emph{Implicit feedback}, such as clicks, likes and purchases, is the  most common  user-item interaction data to train a recommendation model.  In such implicit feedback recommendation, we only have access to positive  samples in the observed user-item interactions, leaving the rest unlabeled. And the goal of a recommender system is to model user preferences from the positive user-item interaction samples.
Over the past decade, progress has been propelled by representation learning advances, from classic matrix factorization \cite{koren2009matrix} and neural CF models \cite{he2017ncf} to graph-based architectures that propagate high-order relational signals, such as NGCF and LightGCN \cite{he2020lightgcn, wang2019neural} to model the user-item interactions.

%% research gpa
While many methods have been proposed to improve the representation learning in recommender systems, the learning objectives  are much less explored. Pairwise ranking losses, represented by Bayesian Personalized Ranking (BPR) \cite{rendle2009bpr}, have long been widely used as learning objectives in implicit-feedback recommender systems due to their simplicity and effectiveness in optimizing positive–negative item rankings.
AdvInfoNCE \cite{zhang2024empowering} introduces contrastive learning to recommender systems by enhancing the standard InfoNCE objective with hard negative mining, thereby constructing more informative contrastive pairs.
Recent studies \cite{wu2024effectiveness} have shown  the Softmax Loss (SL), which normalizes positive user-item interaction over full items into a multinomial distribution, not only improves recommendation accuracy but also enhances  robustness and fairness.
More recently, Wu \etal~ \cite{wu2023bsl} revealed that optimizing SL is equivalent to performing Distributionally Robust Optimization
(DRO) on the negative items, and proposed Bilateral Softmax Loss (BSL) than improves SL by applying the same negative loss structure to  positive  examples. Pairwise Softmax Loss (PSL) \cite{yang2024psl} enhances SL by substituting the exponential formulation with more suitable activation functions, thereby offering a tighter surrogate for the Discounted Cumulative Gain (DCG) metric and alleviating the impact of false negatives.

However, learning from implicit feedback remains fundamentally constrained by \textit{false negatives}. 
Such false negatives introduce biased gradients, distort ranking margins, and reinforce popularity and exposure biases.
A critical disadvantage of both BLS \cite{wu2023bsl} and PSL \cite{yang2024psl} is that they lack an explicit term or mechanism to specifically detect and deal with false negatives. They rely on implicit re-weighting: BSL relies on the mathematical properties of DRO to implicitly weigh hard samples, and PSL utilizes the saturation of activation functions (e.g., Tanh) to control the weight distribution and limit the gradient impact of outliers. Consequently, both methods face the challenge of distinguishing between valuable hard negative samples and detrimental false negatives. As highlighted in the PSL analysis \cite{yang2024psl}, the exponential structure used in BSL can inadvertently assign disproportionately high weights to false negatives that exhibit high prediction scores, and without an explicit correction term, the model may overfit to this noise.

This paper proposes Correct-and-Weight (CW) loss, a simple and plug-and-play loss function for  implicit-feedback recommendation. CW loss addresses false negatives via two complementary mechanisms:
\begin{enumerate}
\item \textbf{Correction of negative sampling via PU-style decontamination.} We reinterpret the sampled negatives as draws from a user-specific mixture of positives and true negatives and derive a corrected objective that replaces the intractable true-negative distribution with a tractable combination of the observable unlabeled and positive distributions, following the spirit of Positive–Unlabeled learning \cite{hsieh2015pu}. This step decontaminates the negative signal inside a Softmax-style objective without requiring changes to the sampler or expensive density estimation.
\item \textbf{Confidence-based re-weighting that anchors positives and easy negatives.} Rather than emphasizing the hardest predicted negatives, which often contain mislabeled positives, CW loss assigns more weight to reliable pairs (observed positives and confidently low-scored negatives) and down-weights uncertain ones. The resulting band-pass emphasis increases the margin where it is most correctable, improving ranking robustness under label noise.
\end{enumerate}

Conceptually, CW loss can be viewed as a minimalist bridge between robust classification with label noise and ranking-oriented CF. By correcting the loss to reflect a cleaner negative distribution and weighting pairs to focus learning on reliable regions, the method sharpens the positive–negative separation while avoiding overfitting to spurious hard negatives. Notably, CW loss is architecture-agnostic and can be readily incorporated into a broad class of recommendation models, including matrix factorization (MF) \cite{koren2009matrix},  graph neural network (GNN)-based recommenders \cite{he2020lightgcn}, and contrastive learning (CL) based methods \cite{yu2023xsimgcl}, with negligible computational overhead and no changes to existing samplers.

We summarize our contributions as follows:
\begin{itemize}
\item We identify and formalize the false-negative contamination in sampled-Softmax style training for implicit feedback recommendation, and derive a PU-corrected objective that substitutes the true-negative sampling distribution with an unbiased combination of observable distributions.

\item We design a confidence-based re-weighting scheme that provably concentrates learning on reliable, correctable pairwise violations, mitigating the adverse influence of mislabeled hard negatives while preserving discriminative margins.

\item  We demonstrate the effectiveness of our proposed CW loss on representative recommendation models (MF, LightGCN, and contrastive GNN variants) with four large-scale, sparse benchmark datasets.

\item We offer empirical analyses and ablations that quantify (1) the benefit of PU-style correction over standard sampled objectives \cite{wu2024effectiveness, wu2023bsl, yang2024psl}, (2) the effect of hardness targeting on robustness, and (3) the compatibility of CW loss with state-of-the-art self-contrastive enhancements \cite{yu2023xsimgcl, zhang2024empowering}.
\end{itemize}

Overall, our proposed CW loss enhances the alignment between training objectives and ranking-based evaluation metrics in the presence of pervasive implicit-feedback noise. Owing to its simplicity and broad compatibility, it serves as an appealing alternative loss to train modern recommendation systems.

\section{Preliminaries}

\subsection{Problem Setup and Notation}

We consider a typical collaborative filtering (CF) based recommender system setting in which we have a set of users $\mathcal{U}$ and items $\mathcal{I}$, and we have access to a record of the users' past user-item interactions  dataset $\mathcal{D} =\{(u, i)\}_{u\in \mathcal{U}, i\in \mathcal{I}}$, where each pair $(u,i) \in \mathcal{D}$ indicates that user $u$ interacted with item $i$. In particular, we focus on \emph{implicit feedback} user-item interactions  such as purchase, clicks,  and likes. For a given user $u$, we define the set of observed positives as $\mathcal{I}_u = \{ i \in \mathcal{I} : (u,i) \in \mathcal{D} \}$; its complement $\mathcal{I} \setminus \mathcal{I}_u$ is treated as negatives. 
To facilitate the subsequent analysis, we denote $\mathcal{I}_u^-$ as the set of true negatives, that is, the items which user $u$ genuinely dislikes. In the context of implicit-feedback recommendation, however, the exact boundary that characterizes $\mathcal{I}_u^-$ remains fundamentally uncertain. Consequently, for each user $u\in \mathcal{U}$, the only observable data available  consist of $\mathcal{I}$ and $\mathcal{I}_u$, which necessitates developing methods capable of inferring user preferences from such partial and noisy feedback. 

Then given the implicit feedback dataset $\mathcal{D}$, the objective of the recommendation task is to learn a scoring function $r_{ui}=f_{\theta}(u,i) : \mathcal{U} \times \mathcal{I} \to \mathbb{R}$ that accurately quantifies the preference of user $u$ for item $i$, and $\theta$ denotes the learnable parameters of the recommender model. Modern recommender systems predominantly employ an embedding-based methodology, wherein each user $u$ and item $i$ are mapped to $d$-dimensional latent representations, denoted as $\mathbf{e}_u, \mathbf{e}_i \in \mathbb{R}^{d}$, respectively. The model then computes a preference score from these embeddings.  Among various preference scores, the normalized  inner product of user and item  embeddings is a widely adopted and empirically effective choice in recommendation~\cite{chen2023adap}. Following the formulation in~\cite{yang2024psl}, we define the scoring function for each user--item pair $(u,i)$ as
$r_{ui} = \frac{1}{2} \cdot \frac{\mathbf{e}_u \cdot \mathbf{e}_i}{\norm{\mathbf{e}_u} \, \norm{\mathbf{e}_i} }$. Finally, the inferred preference scores are used to recommend to
users a list of items that the users have not experienced yet.

Furthermore, we define the score difference between a negative item $j \in \mathcal{I} \setminus \mathcal{I}_u$  and a positive item $i \in \mathcal{I}_u$ for user $u$ as
$d_{uij} = r_{uj} - r_{ui}$,
which serves as a fundamental ranking objective aimed at optimizing relative ordering of items rather than their absolute score values.

\subsection{Softmax Loss for Implicit Feedback Recommendation} %Loss Functions For Recommendation}

In this work, we adopt the Softmax Loss (SL)~\cite{wu2024effectiveness, wu2023bsl, yang2024psl} as the primary optimization objective to train a recommendation model. The formulation of SL is given by
\begin{align}
    \label{eq:sl}
    \mathcal{L}_{SL} &= \mathbb{E}_{(u,i)\in \mathcal{D}} 
    \left[ -\log \left( \frac{\exp(r_{ui}/\tau)}{\sum_{ j\in \mathcal{I}}\exp(r_{uj}/\tau)} \right) \right] \notag \\
    &= \mathbb{E}_{(u,i)\in \mathcal{D}} 
    \left[ \log \sum_{j\in \mathcal{I}} \exp(d_{uij}/\tau) \right],
\end{align}
where $\tau$ denotes the temperature hyperparameter that controls the sharpness of the softmax distribution. The objective of SL is to increase the relative score of positive user-item pairs $(u,i)$ while simultaneously decreasing that of negative pairs $(u,j)$, thereby optimizing the probability of positive instances over negatives. 

Note that, following the approach in~\cite{wu2023bsl, yang2024psl}, we omit the positive term $\exp(r_{ui}/\tau)$ in the denominator of the SL formulation as shown in Equation (\ref{eq:sl}), as its contribution becomes negligible when the number of negative samples is large. Moreover, recent findings~\cite{yeh2022decoupled} indicate that this modification can improve embedding uniformity and empirically yield slightly better performance.

Typically after training a recommendation model $f_{\theta}(u,i)$, we  recommend to users a list of top
items, based on  the preference scores, the users have not experienced yet. The Discounted Cumulative Gain (DCG)~\cite{jarvelin2017ir} is a widely adopted  evaluation metric for assessing ranking quality in recommendation systems. Specifically, for each user $u$, DCG quantifies the ranking performance by discounting the relevance scores based on item positions. It is formally defined as:
\begin{equation}
    \mathrm{DCG}(u) = \sum_{i \in \mathcal{I}_u} \frac{1}{\log_{2}(1 + \pi_u(i))},
    \label{eq:dcg}
\end{equation}
where $\pi_u(i)$ denotes the rank position of item $i$ in the recommendation list for user $u$.

\vspace{5pt}
{\noindent \bf{Discussions}.} The rationale behind employing SL for  implicit feedback recommendation can be interpreted from two complementary perspectives.  
First,  SL has been theoretically shown to serve as an upper bound of the $-\log \mathrm{DCG}$ objective~\cite{bruch2019analysis, yang2024psl}, 
thereby establishing that optimizing SL is equivalent to optimizing the DCG ranking metric.  
Second, from an optimization standpoint, minimizing SL can be viewed as performing \emph{Distributionally Robust Optimization} (DRO)~\cite{shapiro2017distributionally}, which enhances model robustness under distributional shifts. 

However, this robustness comes at the cost of increased sensitivity to noisy data~\cite{nietert2024outlier, zhai2021doro}, as DRO inherently assigns greater emphasis to hard instances with larger losses, thereby amplifying the influence of potentially noisy samples during training~\cite{wu2023bsl}.
Furthermore, gradient analysis of SL~\cite{yang2024psl} reveals that it implicitly applies an adaptive weighting scheme across all negative--positive pairs, where the gradient weight for each sample $(u,i,j)$ is proportional to $\exp(d_{uij}/\tau)$. While this property can be advantageous for hard example mining~\cite{wu2024effectiveness}, facilitating faster convergence, it also causes false negatives, typically associated with larger $d_{uij}$ values, to receive disproportionately large gradient weights. Consequently, the optimization process of SL may become dominated by such false negatives, ultimately leading to degraded recommendation performance.

\section{Proposed Method}
\label{sec:method}
In this section, we introduce the proposed  Corrected and Weighted (CW) loss to address the  aforementioned challenges. 

\subsection{Method Overview}
To alleviate the  adverse impacts of false negatives inherently present in the Softmax Loss (SL), we propose a dual-perspective framework that addresses this issue through \emph{distribution correction} and \emph{adaptive weighting}. 
First, we aim to approximate the true negative distribution by leveraging both the positive interaction distribution and the overall item distribution for each user, thereby achieving a more faithful estimation of user-specific preference structures. 
Second, we introduce an adaptive weighting mechanism that assigns an importance weight to each sampled instance. This mechanism dynamically emphasizes instances with higher confidence, both positives and negatives, allowing the model to focus on more informative samples while reducing the influence of noisy or ambiguous interactions.

\subsection{Correction For The Sampling of Negatives}
\label{Sec:sampling}

\begin{figure}%[th]
\begin{center}
\includegraphics[width=0.49\textwidth]{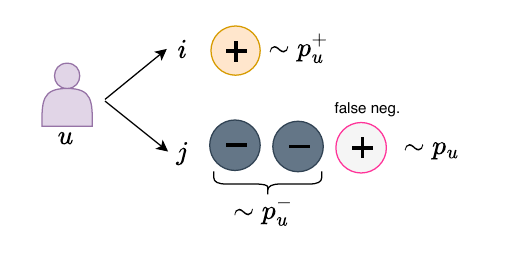} 
\end{center}
\vspace{-10pt}
\caption{Illustration of the false negative problem inherent in implicit feedback recommendation. A negative item $j$ is typically sampled from the general distribution of unlabeled items for a user $u$, denoted as $p_u$. This distribution, however, is a mixture of true negatives (from the underlying distribution $p_u^-$) and  false negatives. The goal of our work is to decontaminate this training signal by formally approximating the true negative distribution $p_u^-$ using the observable distributions $p_u$ and $p_u^-$.}
\vspace{-10pt}
\label{fig:confushion}
\end{figure}

In the context of implicit feedback recommendation, the observed training data typically consist only of positive interactions (e.g., clicks, views, or purchases), while explicit negative labels are absent. As shown in Fig. \ref{fig:confushion}, this asymmetry introduces the well-known issue of false negatives, since unobserved items may be incorrectly treated as negative instances. To address this challenge, we aim to derive a learning objective that more faithfully approximates the ideal unbiased loss. Our strategy is to debias the loss by leveraging the relationship between the overall item distribution $p_u$, the positive item distribution $p_u^+$, and the (unobserved) negative item distribution $p_u^-$. Specifically, adopting the perspective of Positive–Unlabeled (PU) learning \cite{elkan2008learning, du2014analysis, chuang2020debiased}, we express the user-specific sampling distribution as a mixture of positive and negative components:
\begin{equation}
\label{eq:distribution}
p_u(i) = \tau_u^+ p_u^+(i) + \tau_u^- p_u^-(i),
\end{equation}
where $p_u(i)$ denotes the general probability of sampling item $i$ for user $u$. Here, $p_u^+(i)$ and $p_u^-(i)$ represent the conditional distributions over positive and negative items, respectively, and the coefficients $\tau_u^+$ and $\tau_u^-=1-\tau_u^+$ correspond to the prior probabilities that a randomly chosen item for user $u$ belongs to the positive or negative set. Rearranging Eq. \eqref{eq:distribution} yields a formula $p_u^-(i) = \left(p_u(i) - \tau_u^+ p_u^+(i)\right)/\tau_u^- $ for the negative sampling distribution $p_{u}^-$ in terms of two distributions that are tractable since we have samples from $p_u$ and can approximate samples from $p_u^+$ using positive sampling distribution, as is typical in the settings of implicit feedback recommendations. 

Following \cite{yang2024psl}, we rewrite the SL (Equation \ref{eq:sl}) for user $u$ in a pairwise form as:
\begin{equation}
\label{eq:pair_form_SL}
\mathcal{L}_{SL} (u)=\mathbb{E}_{ i \sim p_u^+, j\sim p_u } \left [ \log \left ( \frac{Q}{N}\sum_j \exp(d_{uij}/\tau) \right ) \right ],
\end{equation}
where we introduce a weighting parameter Q for the analysis. When the number $N$ of negative items is finite, we take $Q=N$ in agreement with the standard softmax loss. A fundamental limitation of $\mathcal{L}_{SL}(u)$ lies in its sampling procedure: for each user $u$, the index $j$ is drawn from the distribution $p_u$ over the entire item space $\mathcal{I}$, rather than from the restricted distribution $p_u^-$ defined on the set of true negative items $\mathcal{I}_u^-$. This mismatch implies that when $j$ is sampled from $p_u$, it belongs to the set of positive items with probability $\tau_u^+$. As a result, the effectiveness of negative sampling is diluted, which in turn can introduce bias into the learning process. 

One solution to this problem is that we can replace $p_u^-$ with $p_u$ and $p_u^+$ according to Eq. \eqref{eq:distribution}.

\noindent \textbf{Lemma 1.} For fixed $Q$ and $N \rightarrow \infty$, it holds that
\begin{align}
\mathbb{E}_{i \sim p_u^+, j\sim p_u} \left [ \log \left ( \frac{Q}{N}\sum_j \exp(d_{uij}/\tau) \right ) \right ] \label{eq:debiased-1}\\
\rightarrow \mathbb{E}_{i \sim p_u^+} \bigg[
  \log \Bigg(
    \frac{Q}{\tau^-}
    \Big(
      \mathbb{E}_{j \sim p_u}\big[\exp(d_{uij}/\tau)\big] \notag \\
      - \tau^+ \mathbb{E}_{k \sim p_u^+}\big[\exp(d_{uik}/\tau)\big]
    \Big)
  \Bigg)
\bigg]. \label{eq:debiased-2}
\end{align}

\noindent \emph{Proof.} Since the softmax loss is bounded, we can apply the Dominated Convergence Theorem (DCT) to complete the proof. 
\begin{align}
      &\lim_{N \to \infty} \mathbb{E} \left [  \log \left ( \frac{Q}{N}\sum_{j=1}^{N} \exp(d_{uij}/\tau) \right ) \right ] \notag \\ 
    = &\mathbb{E} \left [ \lim_{N \to \infty} \log \left ( \frac{Q}{N}\sum_{j=1}^{N} \exp(d_{uij}/\tau) \right ) \right ] \mathrm{(DCT)} \notag \\
    = &\mathbb{E} \left [ \log \left ( Q \mathbb{E}_{j \sim p_u^-} [ \exp (d_{uij}/\tau)] \right )  \right ].
\label{eq:debiase-proof}
\end{align}
Since $p_u^-(i) = (p_u(i) - \tau_u^+ p_u^+(i))/\tau_u^- $ and by the linearity of the expectation, we have 
\begin{align}
      &\mathbb{E}_{j \sim p_u^-} [ \exp (d_{uij}/\tau)] \notag \\ 
    = &\frac{1}{\tau^-}\left(\mathbb{E}_{j\sim p_u}[\exp(d_{uij}/\tau)] - \tau^+\mathbb{E}_{k\sim p_u^+}[\exp (d_{uik}/\tau)]\right)
\end{align}
which completes the proof. 

This limiting objective still samples items $j$ from $p_u$, but corrects for that with additional positive items $v$. The empirical estimate in Eq. \eqref{eq:debiased-2} is much easier to compute that the straightforward objective as shown in Eq. (\ref{eq:debiased-1}). With $N$ items $\{j_s\}_{s=1}^N$ from $p_u$ and $M$ items $\{k_s\}_{s=1}^M$ from $p_u^+$, we can estimate the expectation in formula (\ref{eq:debiased-2}).

%%%%%%%%%%%%%%%%%%%%%%%%%%%%%%%%%%%%%%%%%%%%%%%%%%%%%%%%%%%%%%%%%%%%%%%%%%%%%%%%%%%%%%%%%%%%%%%%%%%%%%%%%%%%%%%
\subsection{Confidence-Aware Weight Adjustment}
\label{Sec:weighting}

In addition to considering the sampling distribution of items, another essential strategy for mitigating false negatives in implicit-feedback recommendation systems pertains to the post-sampling treatment of items. A widely adopted approach involves assigning adaptive weights to sampled items to better capture their relative significance. In the domain of computer vision, Robinson \etal ~\cite{robinson2021contrastive} emphasizes the role of hard negatives in contrastive learning by assigning higher weights to negatives exhibiting greater similarity to the anchor instance. Analogously, in recommender systems, Shi ~\etal \cite{shi2023theories} applies a similar principle by allocating higher weights to negatives associated with higher preference scores. However, \cite{wu2023understanding} recently highlighted that negative instances should be assigned balanced and reasonable weights, cautioning against excessive emphasis on highly similar negatives. This concern is particularly pertinent in recommender systems, where the issue of false negatives is especially pronounced. Overemphasizing negatives with high similarity risks misclassifying true positives as hard negatives, thereby misleading the optimization process and further exacerbating model degradation.

\begin{figure}%[th]
\begin{center}
\includegraphics[width=0.49\textwidth]{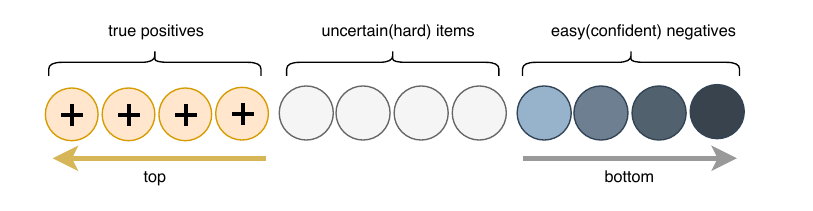} 
\end{center}
\vspace{-10pt}
\caption{The ranking objective of our proposed method. The model is trained to anchor the ranked list by pushing true positives to the top and confident negatives to the bottom. Consequently, uncertain items, which include potential false negatives, are implicitly positioned in the middle ranks.}
% \vspace{-10pt}
\label{fig:weights_}
\end{figure}

In this paper, to further address the issue of false negatives, we introduce an alternative perspective: rather than emphasizing hard negatives, we advocate focusing on positive items and confidently identified (i.e., easy) negatives. This perspective enables recommendation models to generate ranking lists that are both more reliable and better aligned with users’ true preferences. In implicit-feedback scenarios, only positive interactions are explicitly observed. As training proceeds, the model can progressively identify a subset of items that can be confidently regarded as easy negatives. Our approach leverages these two trustworthy categories, positives and easy negatives, to iteratively refine the ranking space. Specifically, these items serve as anchors that guide uncertain (hard) items toward their appropriate positions in the ranking list, as illustrated in Fig.~\ref{fig:weights_}. Based on this analysis, we define the following weighting function:
\begin{equation}
\label{eq:weight}
    w_{uij} = \exp(\beta(r_{ui} - r_{uj})), 
\end{equation}
where $\beta$ denotes a tunable hyperparameter controlling the sensitivity of the weight to score differences.

We adopt the pairwise formulation of the Softmax Loss (SL) as our recommendation objective~\cite{yang2025breaking, yang2024psl}. Incorporating the proposed weight, the new loss function is expressed as:
\begin{eqnarray}
    \label{eq:loss_w}
    \mathcal{L}_W(u) =  \sum_{i \in \mathcal{I}_u} \log \left( 
        \sum_{j \in \mathcal{I}} w_{uij} \, \sigma(d_{uij})^{1/\tau} \right),
\end{eqnarray}
where $\sigma(\cdot)$ can represent one of several activation functions, including $\exp$, $\mathrm{ReLU}$, $\mathrm{tanh}$, and $\mathrm{atan}$~\cite{yang2024psl}. In this work, we employ $\mathrm{ReLU}$ as the activation function. Here, $d_{uij} = r_{uj} - r_{ui}$ denotes the score difference between a negative and a positive item, and $\tau$ is a temperature hyperparameter controlling the smoothness of the distribution.

From the loss function in Eq.~\eqref{eq:loss_w}, it can be observed that the proposed weighting scheme appears counter-intuitive at first glance, as it assigns greater emphasis to easy samples—contrary to the strategies adopted in prior work~\cite{shi2023theories}. However, we argue that easy samples are generally associated with higher confidence, which helps alleviate the false negative problem. When considering Eq. (\ref{eq:loss_w}) in its entirety, the loss exhibits several desirable properties for implicit-feedback recommendation.

\begin{figure}%[h]
\begin{center}
\includegraphics[width=0.30\textwidth]{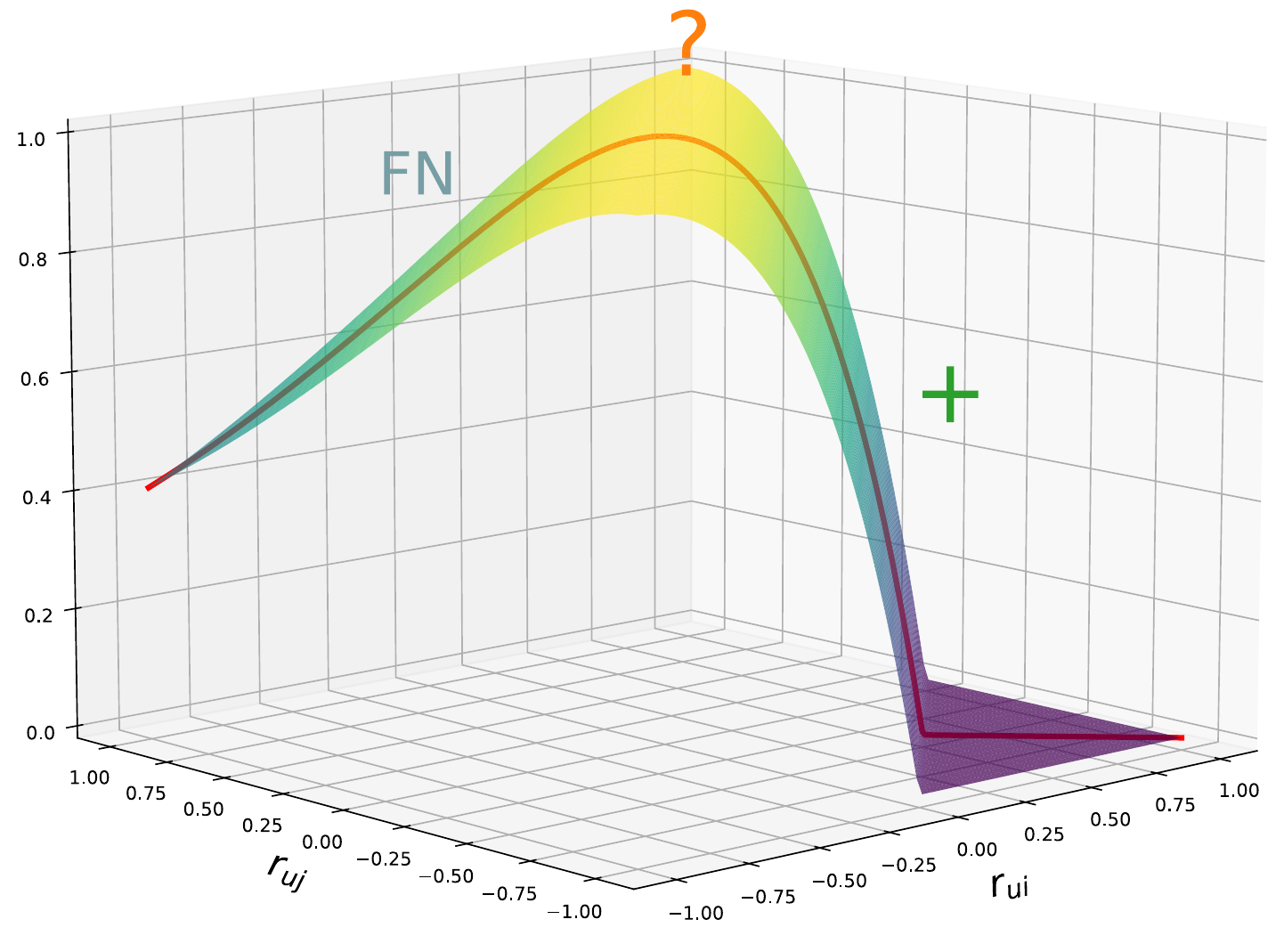} 
\end{center}
\vspace{-10pt}
\caption{The loss curve with respect to $r_{ui}$ and $r_{uj}$. The region labeled FN denotes the area where false negatives are likely to occur. The red ``?'' region represents uncertain instances whose labels or confidence are ambiguous. The + region corresponds to the confident zone where the recommender system has been effectively optimized. } 
% \vspace{-10pt}
\label{fig:curve_fn}
\end{figure}

As illustrated in Fig.~\ref{fig:curve_fn}, when plotting the loss with respect to $r_{ui} $ and $ r_{uj}$, the curve can be divided into three distinct regions: (1) the false negative region, (2) the uncertain region, and (3) the confident region representing well-predicted positives over negatives. Our proposed loss $\mathcal{L}_W$ places greater emphasis on uncertain instances that yield larger losses, thereby encouraging the model to refine its predictions in these ambiguous areas. Unlike previous approaches, hard items in our formulation do not produce disproportionately large gradients; instead, meaningful updates are primarily derived from easy, confidently identified samples that provide stable optimization signals. Furthermore, in the region corresponding to potential false negatives, the loss value decreases adaptively, ensuring that the model places less emphasis on mislabeled instances and thus mitigates the adverse effects of false negatives during training.

\noindent \textbf{Theoretical analysis.}
We theoretically establish that the proposed loss function, denoted as $\mathcal{L}_W$, serves as an upper bound of $-\log \mathrm{DCG}(u)$. Consequently, minimizing $\mathcal{L}_W$ effectively corresponds to maximizing the ranking metric $\mathrm{DCG}$. Based on Eq.~\ref{eq:loss_w} and Eq.~\ref{eq:weight}, the loss $\mathcal{L}_W$ can be reformulated as 
\begin{equation}
    \mathcal{L}_W(u) =  \sum_{i \in \mathcal{I}_u} 
    \log \left(
        \sum_{j \in \mathcal{I}} 
        \exp(-\beta d_{uij}) \, 
        \sigma(d_{uij})^{1/\tau} 
    \right).
\end{equation}

In the limiting case where $\beta \to 0^+$, the proposed loss reduces to $\mathcal{L}_{\mathrm{PSL}}$~\cite{yang2024psl}, which has been rigorously proven to constitute an upper bound of $-\log \mathrm{DCG}$. Furthermore, when $d_{uij} < 0$, it holds that $\delta(d_{uij}) \le \exp(-\beta d_{uij}) \sigma(d_{uij})$, where $\delta(\cdot)$ denotes the Heaviside step function, defined as $\delta(x) = 1$ for $x \ge 0$ and $\delta(x) = 0$ otherwise. Conversely, for $d_{uij} \ge 0$, we have $\exp(-\beta d_{uij}) \sigma(d_{uij}) \le \sigma(d_{uij})$. Moreover, when $d_{uij} \ge 0$ and $\beta \le \frac{\ln{\sigma(d_{uij})}}{d_{uij}}$, it follows that $\exp(-\beta d_{uij}) \sigma(d_{uij}) \ge \delta(d_{uij}) = 1$. We can conclude that when $0 < \beta \le \frac{\ln{\sigma(d_{uij})}}{d_{uij}}$, the proposed loss $\mathcal{L}_{\mathrm{W}}$ is an upper bound of $\delta(d_{uij})$. Due to $\log_{2}(1 + \pi_u(i)) \leq \pi_u(i)$ and Jensen’s inequality \cite{jensen1906fonctions}, we have the following inequality
%These properties collectively substantiate that $\mathcal{L}_W$ provides a smooth and differentiable upper bound approximation to $-\log \mathrm{DCG}$, thereby enabling efficient gradient-based optimization. 
\begin{align}
    -\log \mathrm{DCG}(u) + \log |\mathcal{I}_u|
&\leq -\log \left( \frac{1}{|\mathcal{I}_u|} \sum_{i \in \mathcal{I}_u} \frac{1}{\pi_u(i)} \right) \notag \\ 
&\leq \frac{1}{|\mathcal{I}_u|} \sum_{i \in \mathcal{I}_u} \log \pi_u(i).
\end{align}
The ranking position $\pi_u(i)$ can be mathematically formulated as 
\begin{equation}
    \pi_u(i) = \sum_{j \in \mathcal{I}} \mathbb{I}\big(f(u, j) \ge f(u, i)\big)
= \sum_{j \in \mathcal{I}} \delta(d_{uij})
\end{equation}

So we can have the following conclusion:
\begin{align}
    -\log \mathrm{DCG}(u) \le \log \sum_{j} \delta(d_{uij}) 
\end{align}

So our proposed loss $\mathcal{L}_W$ servers as an upper bound of $-\log \mathrm{DCG}(u)$, which means minimizing the loss equals maximizing the DCG ranking metric.

This weighting formulation is designed to satisfy three key criteria as follows:
\begin{itemize}
    \item Positive-aware hardness: the difficulty of a negative sample is defined relative to its corresponding positive.
    \item Negative correlation with positive scores: the perceived hardness of a negative decreases as the predicted score of the associated positive increases, ensuring consistency in comparative ranking.
    \item Adjustability via hyperparameter $\beta$, the level of emphasis placed on sample hardness can be flexibly controlled, allowing practitioners to balance between stability and discrimination.
\end{itemize}
In this way, the proposed scheme ensures that the model learns from reliable signals while maintaining adaptability to different recommendation contexts.

\subsection{Corrected and Weighted Loss}
Combing the distribution correction  discussed in Section \ref{Sec:sampling} and the adaptive weighting discussed in Section \ref{Sec:weighting}, we  formulate the proposed Corrected and Weighted (CW) loss function  as follows:
\begin{align}
    \label{eq:loss_overview}
    \mathcal{L}_{\mathrm{CW}}(u) = \sum_{i \in p_u^+} \log \frac{Q}{\tau_u^-}   \Big[ & \sum_{j \in p_u} \frac{\exp(-\beta d_{uij})}{Z} \sigma(d_{uij}) \notag \\ 
    & - \frac{\tau_u^+}{M} \sum_{k \in p_u^+} \sigma(d_{uik}) \Big],
\end{align}
where $Z$ is the partition function and $\tau_{u}^+=1-\tau_u^-$ represents the prior probability that a randomly chosen item for user $u$ belongs to the positive set. We provide an ablation study of this parameter in the experimental section. $M$ denotes the number of positive items used to estimate the correction term, and we set $Q$ to the finite $N$ specifying the number of sampled negative items.

\section{Experiments}
In this section, we report our experimental results on four real-world datasets to show the effectiveness of our proposed method.

\subsection{Datasets}

The four benchmark datasets used in our experiments are summarized in Table \ref{tab:statistic}. In dataset preprocessing, following the standard practice in \cite{wang2019neural}, we use 10-core setting \cite{he2016vbpr}, \ie all users and items have at least 10 interactions. We also remove the low-quality interactions, such as those with ratings (if available) lower than 3. After preprocessing, we split the datasets into 80\% training and 20\% test sets. In IID and Noise settings, we further randomly split a 10\% validation set from training set for hyperparameter tuning.
The details of datasets are as follows:
\begin{itemize}%[topsep=3pt,leftmargin=10pt,itemsep=5pt]
    \item \textbf{Health / Electronic / Book \cite{he2016ups, mcauley2015image}:} These datasets are collected from the Amazon dataset, a large-scale collection of product reviews from Amazon. The 2014 version of the Amazon dataset contains 142.8 million reviews spanning May 1996 to July 2014.
    \item \textbf{Gowalla \cite{cho2011friendship}:} The Gowalla dataset is a check-in dataset from the location-based social network Gowalla, including 1 million users, 1 million locations, and 6 million check-ins.
\end{itemize}

% TABLE: DATASET STATISTICS -----------------------------------------
\begin{table}%[tbh]
    \centering
    \caption{Statistics of the datasets we used in this work.}
    \label{tab:statistic}
    \begin{tabular}{l|rrrr}
    \Xhline{1.2pt}
    \multicolumn{1}{c|}{\textbf{Dataset}} & \textbf{\#Users} & \textbf{\#Items} & \textbf{\#Interactions} & \textbf{Density} \bigstrut\\
    \Xhline{1pt}
    Health & 1,974  & 1,200  & 48,189  & 0.02034 \bigstrut[t]\\
    Electronic & 13,455  & 8,360  & 234,521  & 0.00208 \\
    Gowalla & 29,858  & 40,988  & 1,027,464  & 0.00084 \\
    Book  & 135,109  & 115,172  & 4,042,382  & 0.00026 \\
    % MovieLens & 939   & 1,016  & 80,393  & 0.08427 \\
    % Food  & 5,875  & 9,852  & 233,038  & 0.00403 \bigstrut[b]\\
    \Xhline{1.2pt}
    \end{tabular}
\end{table}
% END TABLE: DATASET STATISTICS -------------------------------------

\subsection{Evaluation Metrics}

%This section provides a rigorous description of the evaluation metrics employed in our experiments. 

We adopt the Top-$K$ recommendation setting \cite{liu2009learning}. During evaluation, the positive items present in the training set are masked and excluded from the candidate pool, ensuring that only test-set positives $\mathcal{P}_u$ are considered as ground truth. For each user $u$, we denote the set of correctly recommended items within the Top-$K$ list as $\mathcal{H}_u = \{ i \in \mathcal{P}_u : \pi_u(i) \leq K\}$, where $\pi_u(i)$ represents the rank position of item $i$. The recommendation metrics are formally defined as follows:

\begin{itemize}%[topsep=3pt,leftmargin=10pt,itemsep=5pt]
    \item $ {\mathrm{Recall}@K}$ \cite{fayyaz2020recommendation}: The proportion of ground-truth positives retrieved in the Top-$K$ list, defined as $\mathrm{Recall}@K(u) = |\mathcal{H}_u| / |\mathcal{P}_u|$, with the overall performance given by $\mathrm{Recall}@K = \mathbb{E}_{u \sim \mathcal{U}} [\mathrm{Recall}@K(u)]$.
    
    \item $ {\mathrm{NDCG}@K}$ \cite{jarvelin2017ir}: The Discounted Cumulative Gain at $K$ is defined as $\mathrm{DCG}@K(u) = \sum_{i \in \mathcal{H}_u} 1 / \log_2(1 + \pi_u(i))$. To account for variation in the number of positives $|\mathcal{P}_u|$, $\mathrm{DCG}@K$ is normalized by the ideal $\mathrm{DCG}@K$, yielding $\mathrm{NDCG}@K(u) = \mathrm{DCG}@K(u) / \mathrm{IDCG}@K(u)$, where $\mathrm{IDCG}@K(u) = \sum_{i = 1}^{\min\{K, |\mathcal{P}_u|\}} 1 / \log_2(1 + i)$. The overall score is $\mathrm{NDCG}@K = \mathbb{E}_{u \sim \mathcal{U}} [\mathrm{NDCG}@K(u)]$.

\end{itemize}

\subsection{Recommendation Backbones}

Recommendation backbones (a.k.a. recommendation models) constitute the central components of recommender systems (RS). In the context of this work, a recommendation backbone can be formally characterized as a preference scoring function \( r_{ui}: \mathcal{U} \times \mathcal{I} \rightarrow \mathbb{R} \) parameterized by \(\Theta\), where \(\mathcal{U}\) and \(\mathcal{I}\) denote the user and item spaces, respectively. Evaluating the proposed recommendation loss across diverse backbones is essential to verify its robustness, generalization, and consistency. 

In our experimental framework, we consider three representative recommendation backbones that differ in architectural design and modeling assumptions:

\begin{itemize}%[topsep=3pt,leftmargin=10pt,itemsep=5pt]
    \item \textbf{Matrix Factorization (MF) \cite{koren2009matrix}:} 
    MF is a foundational yet highly effective recommendation approach that decomposes the user--item interaction matrix into latent user and item representations. Many embedding-based models employ MF as the foundational layer for representation learning. In accordance with \cite{wang2019neural} and \cite{yang2024psl}, we set the embedding dimensionality to \(d = 64\) for all experimental configurations.

    \item \textbf{LightGCN \cite{he2020lightgcn}:}  
    LightGCN is a graph-based recommendation model that extends collaborative filtering through graph convolutional operations on the user--item bipartite graph. By removing feature transformation and nonlinear activation components from NGCF \cite{wang2019neural}, LightGCN retains only the essential neighborhood aggregation mechanism. Following the original settings in \cite{he2020lightgcn} and \cite{yang2024psl}, we employ a two-layer architecture.

    \item \textbf{XSimGCL \cite{yu2023xsimgcl}:}  
    XSimGCL introduces a contrastive learning paradigm \cite{jaiswal2020survey, liu2021self} on top of a three-layer LightGCN backbone. It injects random perturbations into the layer-wise embeddings and imposes a contrastive regularization between the final layer and the \(l^{*}\)-th intermediate layer through an auxiliary InfoNCE loss \cite{oord2018representation}. Consistent with \cite{yu2023xsimgcl} and \cite{yang2024psl}, we configure the perturbation magnitude to \(0.1\), set \(l^{*}=1\) (treating the embedding layer as the \(0\)-th layer), use a temperature parameter of \(0.1\), and select the weight of the auxiliary InfoNCE loss from \(\{0.05, 0.1, 0.2\}\).
\end{itemize}

\subsection{Baseline Recommendation Losses}

To comprehensively assess the effectiveness of the proposed loss function, we reproduce several state-of-the-art (SOTA) recommendation losses and determine their optimal hyperparameters via grid search. For loss optimization, we employ the Adam optimizer \cite{kingma2014adam} with two tunable hyperparameters: learning rate (lr) and weight decay (wd). The batch size is fixed at 1024, and the training process is conducted for 200 epochs, within which all compared methods are observed to converge. When negative sampling is required, the number of negative samples is set to \(N = 1000\), except for the MovieLens dataset, where it is reduced to 200 due to the smaller size of its item set. These configurations are consistent with the experimental settings reported in \cite{yang2024psl}. The details of the compared methods and their corresponding hyperparameter search spaces are summarized as follows:

\begin{itemize}%[topsep=3pt,leftmargin=10pt,itemsep=5pt]
\item\textbf{BPR \cite{rendle2009bpr}.}
BPR is a conventional pairwise loss based on Bayesian Maximum Likelihood Estimation (MLE) \cite{casella2024statistical}. The objective of BPR is to learn a partial order among items, \ie positive items should be ranked higher than negative items. Furthermore, BPR is a surrogate loss for AUC metric \cite{rendle2009bpr, silveira2019good}. The score function $r_{ui}$ in BPR is defined as the dot product between user and item embeddings.

\item\textbf{LLPAUC \cite{shi2024lower}.}
LLPAUC is a surrogate loss designed for the lower-left part of AUC. It has been shown to serve as a surrogate loss for Top-$K$ metrics such as Recall@$K$ and Precision@$K$ \cite{shi2024lower,fayyaz2020recommendation}. The score function $r_{ui}$ in LLPAUC is defined as the cosine similarity between user and item embeddings.

\item\textbf{Softmax Loss (SL) \cite{wu2024effectiveness}.}
SL is a SOTA recommendation loss derived from the listwise Maximum Likelihood Estimation (MLE). Beyond explaining the effectiveness of SL from the perspectives of MLE or contrastive learning, it has been demonstrated that SL serves as a DCG surrogate loss. Specifically, SL is an upper bound of $-\log$ DCG \cite{bruch2019analysis, yang2024psl}, ensuring that optimizing SL is consistent with optimizing DCG. In practice, SL introduces a temperature hyperparameter $\tau$ to control the smoothness of the softmax operator. The score function $r_{ui}$ in SL is defined as the cosine similarity between user and item embeddings.

\item\textbf{AdvInfoNCE \cite{zhang2024empowering}.}
AdvInfoNCE is a Distributionally Robust Optimization (DRO) \cite{shapiro2017distributionally}-based modification of SL. It introduces adaptive negative hardness into the pairwise score difference $d_{uij}$ in SL. Though this modification may lead to robustness enhancement, it also enlarges the gap between loss and DCG optimization target, which may lead to suboptimal performance \cite{yang2024psl}. In practical implementation, following the original settings in \cite{zhang2024empowering}, the negative weight is fixed at 64, the adversarial learning is performed every 5 epochs, and the adversarial learning rate is set to $5 \times 10^{-5}$. The score function $r_{ui}$ in AdvInfoNCE is defined as the cosine similarity between user and item embeddings.

\item\textbf{BSL \cite{wu2023bsl}.}
BSL is a DRO-based modification of SL that applies additional DRO to positive instances. It introduces two temperature hyperparameters, $\tau_1$ and $\tau_2$. When $\tau_1 = \tau_2$, BSL is equivalent to SL. The score function $r_{ui}$ in BSL is defined as the cosine similarity between user and item embeddings.

\item\textbf{PSL \cite{yang2024psl}.}
PSL is an NDCG surrogate loss that generalizes SL by substituting the exponential function with a more appropriate activation function. \cite{yang2024psl} proved that PSL establishes a tighter upper bound of $-\log$ DCG than SL, thereby leading to SOTA recommendation performance. Additionally, PSL not only retains the advantages of SL in terms of DRO robustness, but also enhances the noise resistance against false negatives. PSL is also hyperparameter-efficient, requiring only a single temperature hyperparameter $\tau$ to control the smoothness of the gradients. The score function $s_{ui}$ in PSL is defined as \emph{half} the cosine similarity between user and item embeddings.

\item\textbf{SL@$K$ \cite{yang2025breaking}.}
SL@$K$ is a DCG@$K$ surrogate loss proposed in this study. Formally, SL@$K$ is a weighted SL with weight $w_{ui} = \sigma_w(r_{ui} - \beta_{u}^{K})$, where $\beta_{u}^{K}$ is the Top-$K$ quantile of user $u$'s preference scores over all items, and $\sigma_w$ is an activation function (\eg the sigmoid function). Intuitively, the weight $w_{ui}$ is designed to emphasize the importance of Top-$K$ items in the gradients, thereby enhancing Top-$K$ recommendation performance. Compared to the conventional SL, SL@$K$ introduces several hyperparameters, including the temperature hyperparameter $\tau_w$ for the quantile-based weight, the temperature hyperparameter $\tau_d$ for the SL loss term, and the quantile update interval $T_{\beta}$. In practice, $\tau_d$ can be set directly to the optimal temperature $\tau$ of SL. The score function $r_{ui}$ in SL@$K$ is defined as the cosine similarity between user and item embeddings. 
\end{itemize}

% TABLE: IID EXPERIMENTS --------------------------------------------
\begin{table*}[t]
    \centering
    \caption{Comparison of top-20 recommendation performance between our proposed loss (CW) and existing loss functions. The best results are highlighted in bold, and the best baseline performances are underlined. “Imp.” denotes the performance improvement of our loss over the best baseline.}
    \label{tab:cw_versus_sota}
    \small
    \begin{tabularx}{0.98\textwidth}{c l YY YY YY YY}
\toprule
\multirow{2}{*}{\textbf{Backbone}} &
\multicolumn{1}{c}{\multirow{2}{*}{\textbf{Loss}}} &
\multicolumn{2}{c}{\textbf{Health}} &
\multicolumn{2}{c}{\textbf{Electronic}} &
\multicolumn{2}{c}{\textbf{Gowalla}} &
\multicolumn{2}{c}{\textbf{Book}} \\
\cmidrule(lr){3-4}\cmidrule(lr){5-6}\cmidrule(lr){7-8}\cmidrule(lr){9-10}
 & & \textbf{Recall} & \textbf{NDCG} & \textbf{Recall} & \textbf{NDCG}
 & \textbf{Recall} & \textbf{NDCG} & \textbf{Recall} & \textbf{NDCG} \\
\midrule
        \multirow{9}[6]{*}{MF}
            & BPR               & 0.1627  & 0.1234  & 0.0816  & 0.0527  & 0.1355  & 0.1111  & 0.0665  & 0.0453  \bigstrut[t]\\
            & LLPAUC            & 0.1644  & 0.1209  & 0.0821  & 0.0499  & 0.1610  & 0.1189  & 0.1150  & 0.0811  \\
            & SL                & 0.1719  & 0.1261  & 0.0821  & 0.0529  & 0.2064  & 0.1624  & 0.1559  & 0.1210  \\
            & AdvInfoNCE        & 0.1659  & 0.1237  & 0.0829  & 0.0527  & 0.2067  & 0.1627  & 0.1557  & 0.1172  \\
            & BSL               & 0.1719  & 0.1261  & 0.0834  & 0.0530  & 0.2071  & 0.1630  & 0.1563  & 0.1212 \\
            & PSL               & 0.1718  & 0.1268  & 0.0838  & 0.0541  & 0.2089  & 0.1647  & 0.1569  & 0.1227  \\
            & SL@K              & \uline{0.1823} & \uline{0.1390} & \uline{0.0901} & \uline{0.0590} & \uline{0.2121} & \uline{0.1709} & \uline{0.1612} & \uline{0.1269}  \bigstrut[b] \\
            & \textbf{CW(Ours)} & \textbf{0.1908} & \textbf{0.1481} & \textbf{0.0938} & \textbf{0.0618} & \textbf{0.2142} & \textbf{0.1747} & \textbf{0.1634} & \textbf{0.1298} \bigstrut[b]\\
            \cline{2-10}    & \textbf{Imp. \%} & \textbf{+4.71\%} &\textbf{+6.54\%} & \textbf{+4.11\%} & \textbf{+4.76\%} & \textbf{+1.01\%} &\textbf{+2.25\%} &\textbf{+1.35\%}  &\textbf{+2.33\%} \bigstrut\\
        \midrule
        \multirow{9}[6]{*}{LightGCN}
            & BPR               & 0.1618  & 0.1203  & 0.0813  & 0.0524  & 0.1745  & 0.1402  & 0.0984  & 0.0678  \bigstrut[t]\\
            & LLPAUC            & 0.1685  & 0.1207  & 0.0831  & 0.0507  & 0.1616  & 0.1192  & 0.1147  & 0.0810  \\
            & SL                & 0.1691  & 0.1235  & 0.0823  & 0.0526  & 0.2068  & 0.1628  & 0.1567  & 0.1220  \\
            & AdvInfoNCE        & 0.1706  & 0.1264  & 0.0823  & 0.0528  & 0.2066  & 0.1625  & 0.1568  & 0.1177  \\
            & BSL               & 0.1691  & 0.1236  & 0.0823  & 0.0526  & 0.2069  & 0.1628  & 0.1568  & 0.1220  \\
            & PSL               & 0.1701  & 0.1270  & 0.0830  & 0.0536  & 0.2086  & 0.1648  & 0.1575  & 0.1233  \\
            & SL@K   & \uline{0.1783} & \uline{0.1371} & \uline{0.0903} & \uline{0.0591} & \uline{0.2128} & \uline{0.1729} & \uline{0.1625} & \uline{0.1280} \bigstrut[b]\\
            & \textbf{CW(Ours)}    & \textbf{0.1827} & \textbf{0.1454} & \textbf{0.0971} & \textbf{0.0633} & \textbf{0.2154} & \textbf{0.1783} & \textbf{0.1662} & \textbf{0.1306} \bigstrut[b]\\
            \cline{2-10}            & \textbf{Imp. \%} & \textbf{+2.49\%} & \textbf{+6.07\%} & \textbf{+7.46\%} & \textbf{+7.17\%} & \textbf{+1.22\%} & \textbf{+3.14\%} & \textbf{+2.28\%} & \textbf{+2.04\%} \bigstrut\\
        \midrule
        \multirow{9}[6]{*}{XSimGCL}
            & BPR               & 0.1496  & 0.1108  & 0.0777  & 0.0508  & 0.1966  & 0.1570  & 0.1269  & 0.0905  \bigstrut[t]\\
            & LLPAUC            & 0.1519  & 0.1083  & 0.0781  & 0.0481  & 0.1632  & 0.1200  & 0.1363  & 0.1008  \\
            & SL                & 0.1534  & 0.1113  & 0.0772  & 0.0490  & 0.2005  & 0.1570  & 0.1549  & 0.1207  \\
            & AdvInfoNCE        & 0.1499  & 0.1072  & 0.0776  & 0.0489  & 0.2010  & 0.1564  & 0.1568  & 0.1179  \\
            & BSL               & 0.1649  & 0.1201  & 0.0800  & 0.0507  & 0.2037  & 0.1597  & 0.1550  & 0.1207  \\
            & PSL               & 0.1579  & 0.1143  & 0.0801  & 0.0507  & 0.2037  & 0.1593  & 0.1571  & 0.1228  \\
            & SL@K    & \uline{0.1753} & \uline{0.1332} & \uline{0.0869} & \uline{0.0571} & \uline{0.2095} & \uline{0.1717} & \uline{0.1624} & \uline{0.1277} \bigstrut[b]\\
            & \textbf{CW(Ours)}    & \textbf{0.1796} & \textbf{0.1396} & \textbf{0.0903} & \textbf{0.0600} & \textbf{0.2120} & \textbf{0.1792} & \textbf{0.1655} & \textbf{0.1306} \bigstrut[b]\\
            \cline{2-10}            & \textbf{Imp. \%} & \textbf{+2.43\%} & \textbf{+4.85\%} & \textbf{+3.94\%} & \textbf{+5.14\%} & \textbf{+1.21\%} & \textbf{+4.36\%} & \textbf{+1.92\%} & \textbf{+2.27\%} \bigstrut\\
        \bottomrule
    \end{tabularx}
\end{table*}
% END TABLE: IID EXPERIMENTS ----------------------------------------

\subsection{Performance Comparison}
Table~\ref{tab:cw_versus_sota} reports the Top-20 performance of the proposed Correct-and-Weight (CW) loss across four widely-used benchmarks: Gowalla, Book, Health, and Electronic, under three representative model architectures: Matrix Factorization (MF), LightGCN, and XSimGCL. In every configuration, CW demonstrates clear and consistent improvements over existing debiasing losses, highlighting its generality and compatibility with both shallow and graph-based collaborative filtering models.

For the MF backbone, CW achieves the largest absolute gains on the Health and Electronic datasets. Specifically, CW improves NDCG@20 by 6.5\% on Health and by 4.8\% on Electronic compared with the strongest competing baseline. On the Gowalla and Book datasets, CW still exhibits moderate but steady improvements, increasing NDCG@20 by 2.3\% and 2.2\% respectively. These results indicate that even without explicit modeling of user-item graph structure, CW effectively mitigates false-negative bias that often arises in sparse implicit feedback settings.

For LightGCN, CW provides substantial and consistent benefits across all datasets. On Electronic, CW yields the most significant improvement, with Recall@20 and NDCG@20 increasing by approximately 7.5\% and 7.2\%, respectively. On Health, CW outperforms the best baseline by 6.0\% in NDCG@20, while maintaining stable gains on Gowalla (3.1\%) and Book (2.0\%). This consistent improvement demonstrates that integrating loss-level debiasing with message-passing architectures allows the model to learn more reliable item embeddings and propagate less biased signals through the user–item graph.

When combined with the contrastive backbone XSimGCL, CW continues to yield competitive results. Even though XSimGCL already employs a self-supervised objective to enhance representation quality, CW still improves NDCG@20 by 4.9\% on Health and 5.1\% on Electronic, while providing smaller yet consistent gains on Gowalla and Book (both around 2.3\%). This observation suggests that the debiasing effect of CW is complementary to self-supervised regularization and can further reduce ranking bias at the objective level, beyond what augmentation-based contrastive learning achieves.

Overall, the results across all architectures and datasets highlight the robustness of CW. The improvements are not confined to a specific model or data domain, confirming that correcting false negatives at the loss level and assigning adaptive confidence weights to individual instances together produce consistent top-$K$ ranking enhancements. This stability underscores CW’s potential as a plug-and-play replacement for traditional softmax-based ranking losses in implicit feedback recommendation.

\begin{figure*}%[h]
    \centering
    \includegraphics[width=1.00\textwidth]{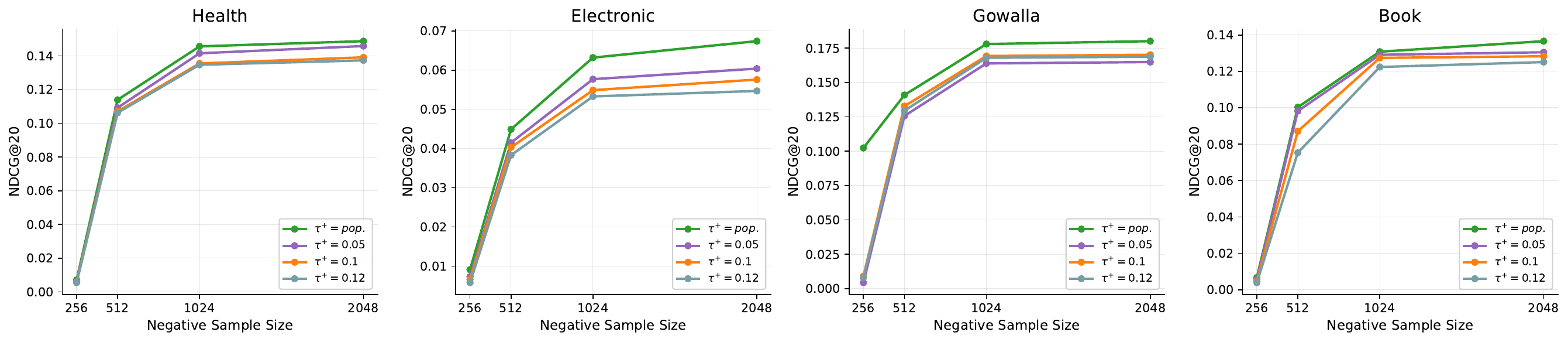}
    \caption{NDCG@20 performance of the proposed loss under varying values of \(\tau^+\) using LightGCN as the recommendation model. The term \textrm{pop.} denotes that \(\tau^+\) is computed based on item popularity.
}
    \label{fig:tau_plus}
\end{figure*}

\subsection{Ablation Study}
To better understand the contribution of each component within CW, we conduct a series of ablation experiments focusing on two main aspects: (1) the correction term that rebalances the negative distribution using a positive-unlabeled (PU) formulation, and (2) the confidence-aware weighting that adaptively scales gradients based on reliability estimates of the training pairs.

\begin{table}[th]
    \centering
    \caption{Different loss factors on four datasets (metrics: NDCG@20, models: LightGCN).}
    \label{tab:loss_c_w}
    \begin{tabular}{l ccccc}
    \Xhline{1.2pt}
    \multicolumn{1}{c }{} & 
    $\mathcal{L}_{\mathrm{SL}}$ & 
    $\mathcal{L}_{\mathrm{PSL}}$ & 
    $\mathcal{L}_{\mathrm{C}}$ & 
    $\mathcal{L}_{\mathrm{W}}$ & 
    $\mathcal{L}_{\mathrm{CW}}$  \bigstrut\\
    \Xhline{1pt}
    Health       & 0.1235 &0.1270 & 0.1243 & 0.1279 & 0.1454 \bigstrut[t]\\
    Electronic   & 0.0526 &0.0536 & 0.0529 & 0.0553 & 0.0633 \\
    Gowalla      & 0.1628 &0.1648 & 0.1632 & 0.1657 & 0.1783 \\
    Book         & 0.1220 &0.1233 & 0.1225 & 0.1261 & 0.1306\\
    % NeuralNDCG   & \uline{0.4338} & 0.4524 & 0.5823 & -\\
    % SL \cite{wu2024effectiveness} & 0.4310 & 0.4552 & \uline{0.6327} & - \\
    % \textbf{SL@5 (Ours)} & \textbf{0.4633} & \textbf{0.4895} & \textbf{0.6412} & - \bigstrut[b]\\
    \hline
    %\textcolor{red}{\textbf{Imp. \%}}   & \textcolor{red}{\textbf{+6.80\%}} & \textcolor{red}{\textbf{+6.55\%}} & \textcolor{red}{\textbf{+1.34\%}} & - \bigstrut\\
    \Xhline{1.2pt}
    \end{tabular}
\end{table}

Table~\ref{tab:loss_c_w} presents the results using the LightGCN backbone with five variants of the loss: the standard softmax loss (denoted $\mathcal{L}_{SL}$), its pairwise form ($\mathcal{L}_{PSL}$), correction-only ($L_{C}$), weighting-only ($\mathcal{L}_{W}$), and the full combination ($\mathcal{L}_{CW}$). Two consistent patterns emerge. First, both $L_{C}$ and $\mathcal{L}_{W}$ outperform $\mathcal{L}_{SL}$ and $\mathcal{L}_{PSL}$ across all datasets, indicating that the correction term successfully removes the contamination from false negatives, while the weighting scheme stabilizes optimization by emphasizing trustworthy positive–negative pairs. Second, when both mechanisms are applied together in $L_{CW}$, the performance improves further, achieving the best results in every setting. For example, on the Electronic dataset, NDCG@20 rises from 0.0526 (with $\mathcal{L}_{SL}$) to 0.0633 (with $\mathcal{\mathcal{L}}_{CW}$), and on Health, from 0.1235 to 0.1454. The additive nature of these gains demonstrates that the two components target distinct but complementary sources of bias.

We further examine the sensitivity of CW to the positive prior $\tau_{u}^{+}$, which specifies the estimated proportion of unobserved items that are actually positive for each user. As shown in Figure~\ref{fig:tau_plus}, CW remains robust under a wide range of $\tau^{+}$ values, specifically between 0.05 and 0.12, and when using a data-driven popularity-based estimate. Several insights can be drawn from this analysis. First, CW exhibits broad stability: performance changes smoothly and modestly as $\tau^{+}$ varies, implying that the method does not require precise prior estimation to work effectively. Second, moderate priors (around 0.1–0.12) tend to perform slightly better than very small ones, suggesting that excessively conservative priors may lead to under-correction. Third, the popularity-informed prior performs comparably or even slightly better than fixed constants, offering a simple, hyperparameter-free way to adapt correction strength to dataset skewness.

\begin{figure}%[h]
\begin{center}
\includegraphics[width=0.4\textwidth]{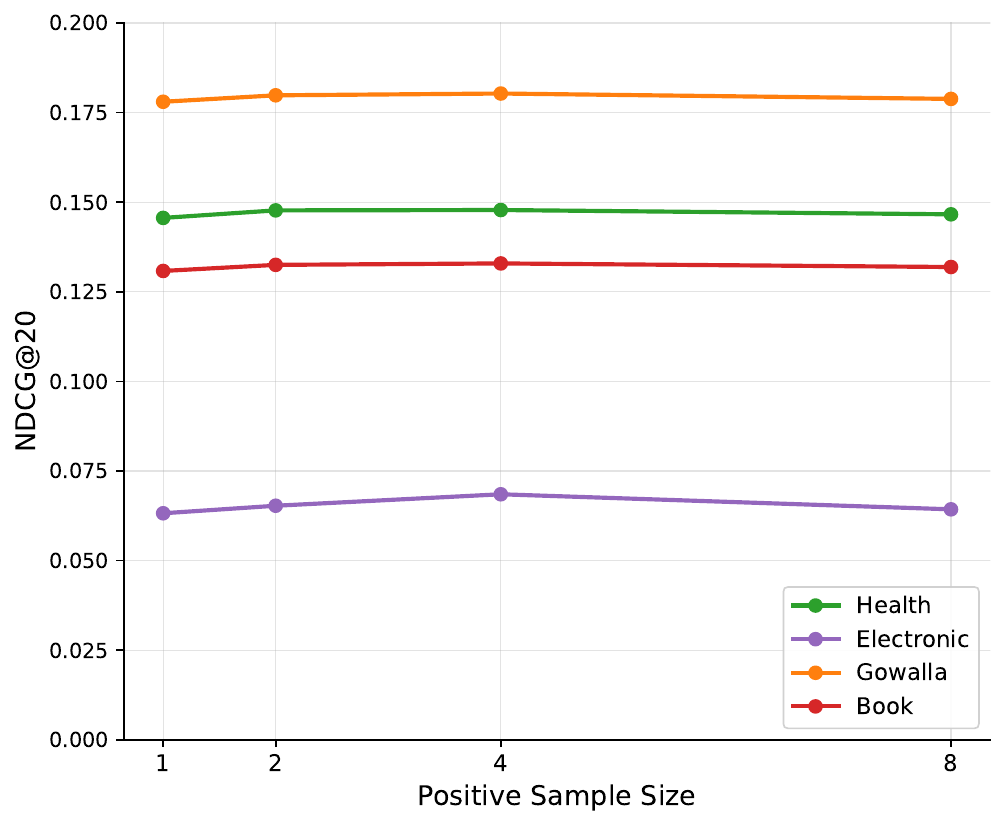} 
\end{center}
\vspace{-10pt}
\caption{Comparison of the proposed method with different numbers of positive samples across four datasets. The results are based on LightGCN model. } 
% \vspace{-10pt}
\label{fig:pos_size}
\end{figure}

\begin{figure*}%[h]
\begin{center}
\includegraphics[width=0.95\textwidth]{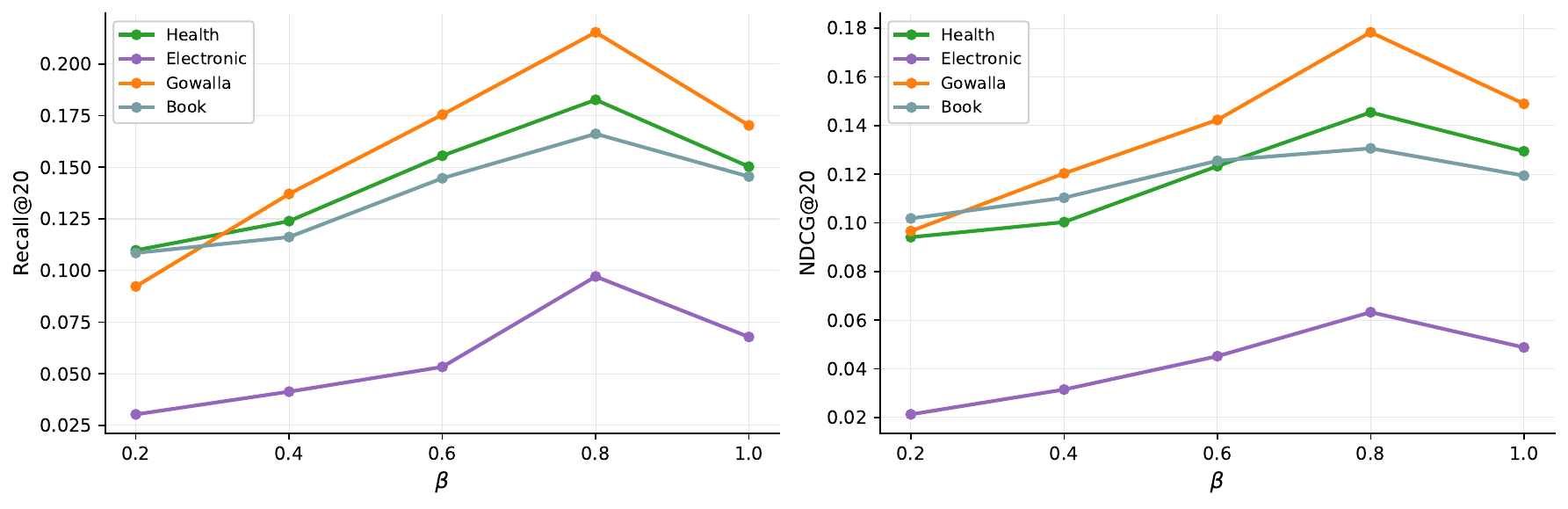} 
\end{center}
\vspace{-10pt}
\caption{Comparison of the proposed method under different values of $\beta$ across four datasets, based on the LightGCN model. } 
% \vspace{-10pt}
\label{fig:weight_beta}
\end{figure*}

We also analyze the effect of the number of positive samples per user used in the corrected objective. As shown in Figure~\ref{fig:pos_size}, the NDCG@20 consistently increases as more positive items are included, with performance saturating after approximately four positives. This finding indicates that CW benefits from richer positive supervision but remains efficient, as a small number of positives per user are sufficient to achieve substantial debiasing effects.

Figure~\ref{fig:weight_beta} presents the sensitivity analysis of the hyperparameter $\beta$ on model performance across four benchmark datasets, utilizing the LightGCN backbone. The behavior of $\beta$ dictates the loss function's focus: as $\beta \to 0$, the weighted component of our proposed loss reduces to the standard PSL loss; conversely, as $\beta$ increases, the model assigns greater importance to hard (uncertain or false) negatives. Consequently, the rationale for optimizing $\beta$ lies in balancing the effective mining of informative hard negatives with the necessity of maintaining robustness against potential noise and outliers within the training data. Empirically, both Recall@20 and NDCG@20 exhibit a consistent upward trend as $\beta$ increases from $0.2$ to $0.8$, indicating that a moderate amplification of hard negative weights enhances the model's capacity to capture latent user preferences and improves overall ranking quality. However, performance degradation is observed when $\beta$ exceeds $0.8$. This decline suggests that excessive emphasis on uncertain negatives may lead to overfitting or the amplification of noise, thereby adversely affecting recommendation accuracy. In terms of dataset-specific performance, \textit{Gowalla} demonstrates the most significant improvement, followed by \textit{Health} and \textit{Book}, while \textit{Electronic} exhibits relatively marginal gains. These results collectively suggest that an intermediate value of $\beta \approx 0.8$ yields the optimal trade-off between recommendation accuracy and model generalization.

In summary, the ablation studies provide strong empirical evidence that (1) the correction term effectively counteracts false-negative contamination, (2) the confidence-aware weighting prioritizes informative samples and enhances optimization stability, and (3) the combination of both mechanisms yields a robust and generalizable loss function. These insights explain the consistent improvements observed in the main experiments and validate CW as a practical, theoretically grounded enhancement to implicit-feedback recommendation models.

\section{Related Work}
Our work is related to recommendation models and objective functions for recommender systems. 

\subsection{Recommendation Models}

The landscape of collaborative filtering (CF) modeling for recommender systems has evolved significantly over the past two decades, with a steady progression from shallow linear methods to highly expressive graph-based architectures. Early research was dominated by Matrix Factorization (MF) based approaches \cite{koren2009matrix}, which established the foundation for representing users and items in a latent space where preference signals are captured by inner products. Extensions of MF introduced increasingly sophisticated mechanisms to better capture the underlying user–item relationships \cite{fiesler2020handbook}. Classic models include SVD \cite{deerwester1990indexing,bell2007modeling}, SVD++ \cite{koren2008factorization}, LRML \cite{tay2018latent}, and NCF \cite{he2017ncf}, each progressively enriching the interaction modeling capacity while preserving computational scalability. These methods laid a theoretical and practical foundation for subsequent innovations by demonstrating how embedding-based representations can compactly encode collaborative signals.

In parallel, the rapid rise of Graph Neural Networks (GNNs) opened new opportunities for CF research. Since CF inherently assumes that users and items are connected through high-order relational structures (e.g., shared preferences or co-consumption patterns), the inductive bias of GNNs aligns naturally with the problem setting \cite{wu2022graph,gao2022graph,kipf2016semi,wang2019neural,dong2021equivalence,wu2022graph_gcm,chen2024macro}. As a result, a new generation of GNN-based CF models emerged, including NGCF \cite{wang2019neural}, LightGCN \cite{he2020lightgcn}, LCF \cite{yu2020graph}, and APDA \cite{zhou2023adaptive}. These approaches systematically exploit higher-order neighborhood aggregation to refine user and item embeddings, thereby achieving superior performance compared to traditional factorization-based techniques.

Building on the structural advances of LightGCN, researchers have also integrated contrastive learning into CF, motivated by its effectiveness in representation learning \cite{liu2021self,oord2018representation}. Contrastive objectives provide self-supervised signals that encourage robust embedding learning through graph-based augmentation. Models such as SGL \cite{wu2021self} and XSimGCL \cite{yu2023xsimgcl} exemplify this line of work, achieving state-of-the-art (SOTA) results by combining GNN architectures with contrastive regularization. Collectively, this trajectory illustrates the transition of CF research from matrix-based latent factorization to graph-based self-supervised paradigms, highlighting a continual pursuit of richer interaction modeling, stronger generalization, and scalability in large-scale recommendation environments.

\subsection{Objective Functions for Recommender Systems} 

While recomendation models have been well studied, the learning objectives for recommender systems are much less explored.
Recommendation losses can be broadly categorized into three canonical classes: pointwise losses \cite{he2017ncf,he2017nfm}, which treat recommendation as a regression or classification task on observed interactions; pairwise losses \cite{rendle2009bpr}, which optimize relative rankings between positive and negative pairs; and Softmax-based losses (SL) \cite{wu2024effectiveness}, which adopt a classification-like view by normalizing positive user-item interaction over full items into a multinomial distribution.

Specifically, owing to their simplicity and effectiveness in optimizing positive–negative item rankings, pairwise ranking losses \cite{rendle2009bpr} have long been widely adopted as learning objectives in implicit-feedback recommender systems. 
Instead of optimizing positive–negative item rankings, Shi \etal~
\cite{shi2024lower} developed LLPAUC, a principled surrogate that approximates the non-differentiable Recall@$K$ metric. 
Building up the standard InfoNCE objective, which samples  a subset of negative instances for each positive instance,  the AdvInfoNCE \cite{zhang2024empowering} adopts contrastive learning  for recommendation via informative hard negative sampling. 

Recently, Softmax-based losses (SL) \cite{wu2024effectiveness} has emerged as particularly effective in practice, partly due to its capacity to leverage full-list information and provide smooth gradients.
Recognizing the strength of SL, recent studies \cite{wu2023bsl,yang2024psl,yang2025breaking} have sought to refine it from various angles. 
Wu et al. \cite{wu2023bsl} demonstrate that SL optimization corresponds to Distributionally Robust Optimization (DRO) on negative samples and introduce Bilateral Softmax Loss (BSL) to  enhance SL by symmetrizing the loss over positive examples.
Yang et al. \cite{yang2024psl} introduce Pairwise Softmax Loss (PSL), an extension of SL that replaces the exponential formulation with alternative activation functions,  and is shown to be a tighter surrogate for Discounted Cumulative Gain (DCG) metric and more robust to false negatives. 
To directly optimize the Top-$K$ ranking metric NDCG@$K$, SoftmaxLoss@$K$ \cite{yang2025breaking} relaxes it into a smooth surrogate loss based on Softmax-based loss.

These efforts highlight an emerging consensus that SL serves as a flexible backbone objective, yet requires careful adaptation to specific challenges such as robustness to false negatives, and better optimizing recommendation performances (\eg ranking metrics).
Different from the efforts \cite{wu2023bsl,yang2024psl,yang2025breaking} to enhance Softmax-based losses for implicit recommendation, the proposed Correct-and-Weight (CW) loss  debias the negative sampling process via approximating the true negative distribution 
using the observable general data distribution  and the positive interaction distribution. Meanwhile, through a dynamic re-weighting mechanism, CW loss encourages larger ranking margins between positive items and confidently predicted negatives, while adaptively reducing penalties on uncertain negatives with a higher likelihood of being false negatives, thereby improving ranking performance.

\section{CONCLUSION and FUTURE WORK}
In this paper, we introduced Correct-and-Weight (CW) loss, a simple yet effective loss function for  implicit feedback recommendation. Unlike existing approaches that rely on complex sampling heuristics or auxiliary modeling assumptions, CW directly corrects the loss itself by decomposing the observed feedback into positive and unlabeled components, and then weighting the training instances according to their reliability. This design simultaneously addresses two long-standing challenges in implicit recommendation: the false-negative contamination in unobserved items and the instability caused by treating all instances with equal importance.

Extensive experiments on four benchmark datasets and multiple backbones, including MF, LightGCN, and XSimGCL, demonstrate that CW consistently outperforms representative state-of-the-art losses. The empirical results validate that (1) correction effectively rebalances the label noise distribution, (2) confidence weighting stabilizes optimization by focusing learning on trustworthy regions, and (3) their combination generalizes across architectures and data regimes. The ablation analyses further confirm that CW remains robust to variations in the positive prior, negative sampling size, and the number of observed positives per user, indicating its practicality for large-scale real-world deployment.

Beyond its empirical performance, CW contributes a conceptually unifying perspective on implicit feedback debiasing. Rather than altering the model structure or introducing external bias estimators, CW reinterprets the learning objective as an estimation-corrected risk minimization problem under positive-unlabeled data assumptions. This connection provides theoretical clarity and highlights that loss-level corrections can serve as a lightweight yet powerful alternative to model-based debiasing strategies.

Although CW already achieves strong results, several directions remain open for future work. First, the current formulation assumes that the prior of latent positives is independent across users, which may not hold in domains with strong user heterogeneity. Future extensions could explore personalized or adaptive priors estimated from user interaction histories or side information. Second, while our analysis focuses on static recommendation, incorporating CW into dynamic or session-based recommendation frameworks could further reveal its potential under evolving user preferences. Third, the confidence weighting mechanism could be extended with uncertainty-aware or Bayesian modeling to provide theoretical guarantees on convergence and generalization. Finally, combining CW with reinforcement or bandit-based recommendation paradigms might offer new insights into integrating debiasing with long-term reward optimization.

\bibliographystyle{IEEEtran}

\bibliography{refs}

\end{document}